\documentclass[10pt,twocolumn,letterpaper]{article} 

\usepackage{avss}
\usepackage{times}
\usepackage{epsfig}
\usepackage{graphicx}
\usepackage{amsmath}
\usepackage{amssymb}
\usepackage{tablefootnote}
\usepackage{multirow}

\avssfinalcopy % *** Uncomment this line for the final submission

\def\avssPaperID{37} % *** Enter the AVSS Paper ID here

\ifavssfinal\fi
\begin{document}

%%%%%%%%% TITLE
\title{XLSR-Kanformer: A KAN-Intergrated model for Synthetic Speech Detection}

\author{Phuong Tuan Dat$^*$\\
Hanoi University of Science and Technology\\
Vietnam\\
{\tt\small phuongtuandat2915@gmail.com}
\and
Tran Huy Dat\\
Institute for Infocomm Research (I\textsuperscript{2}R)\\
A*STAR, Singapore\\
{\tt\small hdtran@i2r.a-star.edu.sg}
}

\maketitle
% \thispagestyle{empty}

%%%%%%%%% ABSTRACT
\begin{abstract}
   Recent advancements in speech synthesis technologies have led to increasingly sophisticated spoofing attacks, posing significant challenges for automatic speaker verification systems. While systems based on self-supervised learning (SSL) models, particularly the XLSR-Conformer architecture, have demonstrated remarkable performance in synthetic speech detection, there remains room for architectural improvements. In this paper, we propose a novel approach that replaces the traditional Multi-Layer Perceptron (MLP) in the XLSR-Conformer model with a Kolmogorov-Arnold Network (KAN), a powerful universal approximator based on the Kolmogorov-Arnold representation theorem. Our experimental results on ASVspoof2021 demonstrate that the integration of KAN to XLSR-Conformer model can improve the performance by 60.55\% relatively in Equal Error Rate (EER) LA and DF sets, further achieving 0.70\% EER on the 21LA set. Besides, the proposed replacement is also robust to various SSL architectures. These findings suggest that incorporating KAN into SSL-based models is a promising direction for advances in synthetic speech detection. \let\thefootnote\relax\footnotetext{$^*$Work done during my internship at A$^*$STAR, Singapore.}
\end{abstract}

%%%%%%%%% BODY TEXT
\section{Introduction}
\label{sec:intro}

Automatic Speaker Verification (ASV) systems play a crucial role in identity authentication by analyzing unique vocal characteristics of individuals. These systems have been widely adopted in real-world security applications, ranging from access control to financial transaction verification. Over the years, ASV technology has seen significant advancements \cite{8461375} \cite{zhang2022mfaconformermultiscalefeatureaggregation}, not only in improving speaker recognition accuracy but also in addressing one of its most critical challenges - Synthetic Speech Detection (SSD). Despite these improvements, the rapid progress in synthetic speech generation techniques, such as Text-to-Speech (TTS) and Voice Conversion (VC), has posed an increasing threat to ASV systems. Attackers can exploit these technologies to generate highly realistic artificial voices that bypass security measures, making ASV systems vulnerable to various malicious attacks. This growing concern highlights the urgent need for more robust and reliable countermeasures to enhance the security and resilience of ASV systems against synthetic speech threats.

Recent studies in speech processing have increasingly leveraged the power of self-supervised learning (SSL) models \cite{truong24b_interspeech} \cite{rosello23_interspeech} \cite{Vaessen_2022}. These models are trained on vast amounts of unlabeled data, enabling them to learn powerful and generalizable representations. Consequently, SSL models can be fine-tuned with labeled data to adapt to various downstream tasks, such as audio classification, speech enhancement, and spoof detection. In particular, \cite{rosello23_interspeech} utilizes an SSL model as a feature extractor to derive meaningful representations from raw waveform inputs. These extracted features are then processed by a Conformer-based architecture, which learns to classify the input effectively. In this paper, we propose a novel architecture to replace the standard Transformer or Conformer, aiming to enhance the learning of features extracted from SSL models by leveraging Kolmogorov-Arnold Networks (KANs). 

Multi-Layer Perceptrons (MLPs) have long been a fundamental component of deep learning architectures, serving as essential building blocks for models like Transformers and Convolutional Neural Networks (CNNs). Despite their widespread adoption, MLPs exhibit limitations when handling high-dimensional data, such as speech signals represented as raw waveforms, spectrograms, or deep embeddings. Their inability to inherently capture complex functional mappings often leads to scalability and efficiency issues, potentially resulting in suboptimal performance in speech-related tasks.

Recently, Kolmogorov-Arnold Networks (KANs) \cite{liu2024kankolmogorovarnoldnetworks} have emerged as a novel alternative to MLPs, drawing inspiration from the Kolmogorov-Arnold representation theorem. Unlike traditional MLPs, which rely on fixed activation functions, KANs employ learnable univariate activation functions, offering greater flexibility and improved function approximation capabilities \cite{SCHMIDTHIEBER2021119}. These properties make KANs particularly well-suited for high-dimensional data, including speech representations extracted from self-supervised learning (SSL) models. Furthermore, a critical challenge when fine-tuning SSL models is catastrophic forgetting, where a model loses previously learned knowledge when adapting to new data. KANs, with their learnable activation functions and adaptive nature, have demonstrated resilience against catastrophic forgetting, making them a promising alternative to MLPs in continual learning scenarios. Motivated by these advantages, we propose leveraging KANs in place of standard MLPs to enhance feature learning in SSL-based speech processing models, improving both robustness and generalization.

We employ the pre-trained XLSR model \cite{conneau2020unsupervisedcrosslingualrepresentationlearning} and our proposed Kanformer encoders - a Conformer architecture enhanced with KAN to improve the detection of synthetic speech. Our model achieves the new state-of-the-art on the ASVSpoof2021 DF set while remains on par with other competitive systems. The subsequent sections in this paper are organized as follows: Section \ref{sec:theory} outlines the theoretical foundation of KANs and their extensions, while Section \ref{sec:method} details the methodologies utilized in this study. In Section \ref{sec:experiments}, we present the experimental setup and results, comparing our proposed model against state-of-the-art (SOTA) approaches. Finally, Section \ref{sec:conclusion} summarizes our key contributions.

%------------------------------------------------------------------------ 
\section{Preliminary}
\label{sec:theory}

% The following section outlines all methods utilized in this research, including KANs, GR-KAN, the baseline model architecture, and our proposed new model architecture.

\subsection{Multi-Layer Perceptrons (MLPs)}

A Multi-Layer Perceptron (MLP) is a type of fully connected feedforward neural network composed of multiple layers of neurons. In this architecture, each neuron establishes connections with all nodes in the subsequent layer, applying a nonlinear activation function to the weighted sum of its inputs.  

MLPs are grounded in the Universal Approximation Theorem \cite{hornik1989multilayer}, which asserts that a feedforward network with a single hidden layer and a finite number of neurons can approximate any continuous function on compact subsets of \(\mathbb{R}^n\), provided that suitable activation functions are used.

\subsection{Kolmogorov-Arnold networks (KANs)}
\label{ssec:kan}

KAN is based on the Kolmogorov-Arnold representation theorem, which states that any continuous multivariate function can be represented as a sum of univariate functions \cite{liu2024kankolmogorovarnoldnetworks}. More specifically, each KAN layer \( L \) is constructed using a combination of learnable univariate functions \( \phi \), as shown in Equation \ref{eq:kan}, where \( d_{in} \) and \( d_{out} \) represent the input and output dimensions, respectively. These univariate functions serve as a means to approximate continuous functions.

\begin{equation}
\label{eq:kan}
    L(\textbf{x}) = \Phi \circ x = [\sum_{i=1}^{d_{in}} \phi_{i,1}(x_{i}) ,. . .  
  ,\sum_{i=1}^{d_{in}} \phi_{i,d_{out}}(x_{i})],
\end{equation}
where \(\Phi\) captures the combined mapping across the input dimensions. 

In practical implementations, a KAN layer \( L \) is approximated using Equation 2 to enhance computational efficiency, where \( w_1 \) and \( w_2 \) are learnable scalars, and \( b \) is a basis function (e.g., SiLU) that functions similarly to residual connections. The \(spline\) refers to a B-spline function, which constructs smooth curves by combining multiple low-degree polynomial segments, effectively representing high-degree polynomial functions.

\begin{equation}
\label{eq:phi}
    \phi(x) = w_{b}\cdot b(x) + w_{s}\cdot spline(x),
\end{equation} 

\subsection{Chebyshev Kolmogorov-Arnold networks}
\label{ssec:chebykan}
Chebyshev Kolmogorov-Arnold networks (ChebyKAN) \cite{ss2024chebyshevpolynomialbasedkolmogorovarnoldnetworks} is developed based on the Kolmogorov-Arnold  representation theorem and incorporate with the powerful approximation capabilities of Chebyshev polynomials. At the heart of this framework are Chebyshev polynomials, a set of orthogonal polynomials defined over the range \([-1, 1]\), making them highly effective for function approximation. The Chebyshev KAN layer, which leverages these polynomials, provides an innovative substitute for traditional B-splines, overcoming their constraints in terms of both performance and usability. Our enhanced Conformer block integrates ChebyKAN transformations, enhancing the architecture's flexibility and adaptability in approximating high-dimensional feature representations. To validate its effectiveness, we conducted ablation experiments in Section \ref{sec:ablation}, demonstrating the feasibility of this approach.

The feature embedding of shape \((B, C, H)\) is reshaped to  \((N, H)\) where \(N = B \times C\), and processed through the Chebyshev KAN layer, where Chebyshev polynomials \(T \in \mathbb{R}^{N \times H \times (\text{degree}+1)}\) are calculated recursively as follows: \(T_0(x) = 1\), \(T_1(x) = x\), and \(T_m(x) = 2xT_{m-1}(x) - T_{m-2}(x)\) for \(m \geq 2\). A trainable tensor of Chebyshev coefficients \(C \in \mathbb{R}^{H \times \text{output\_dim} \times (\text{degree}+1)}\) is introduced, acting as learnable parameters. The output of the layer \(y \in \mathbb{R}^{N \times \text{output\_dim}}\) is computed using Einstein summation, expressed as:

\begin{equation}
\label{eq:Chebyshev}
    y_{no} = \sum_{i=1}^{H} \sum_{j=0}^{\text{degree}} T_{nij} \cdot C_{ioj}
\end{equation} 

Here, \(n\), \(i\), \(o\), and \(j\) correspond to the indices for patches, input dimensions, output dimensions, and polynomial degrees, respectively.

\subsection{Kolmogorov-Arnold Convolution}

By incorporating the adaptable non-linear activation functions from KANs into the Convolutional Neural Networks (CNN) architecture, \cite{bodner2024convolutionalkolmogorovarnoldnetworks} introduces Kolmogorov-Arnold Convolution - a novel alternative to the standard CNN. Kolmogorov-Arnold Convolutions can be described as follows: the convolution kernel is composed of a collection of univariate non-linear functions. For an input image \(y\), where \(y \in \mathbb{R}^{c \times n \times m}\), \(c\) represents the number of channels, and \(n, m\) denote the height and width of the image respectively, the KAN-based convolution with a kernel size of \(k\) is mathematically expressed as:

\begin{equation}
\label{eq:kan_conv}
    x_{ij} = \sum_{d=1}^c \sum_{a=0}^{k-1} \sum_{b=0}^{k-1} \varphi_{a,b,d}(y_{d,i+a,j+b}),
\end{equation} 
where \(i = 1, \ldots, n-k+1\) and \(j = 1, \ldots, m-k+1\).

Here, each \(\varphi\) represents a univariate non-linear learnable function with its own set of trainable parameters as mentioned in \ref{ssec:kan}. In this paper, we also aplly ChebyKAN into the Kolmogorov-Arnold architecture to enhance the model's performance, as discussed in \ref{ssec:chebykan}.

\section{Proposed Methodology}
\label{sec:method}
\subsection{Baseline Model Architectures}
\label{ssec:xlsr}
\begin{figure}[t]
\begin{center}
   \includegraphics[width=0.8\linewidth]{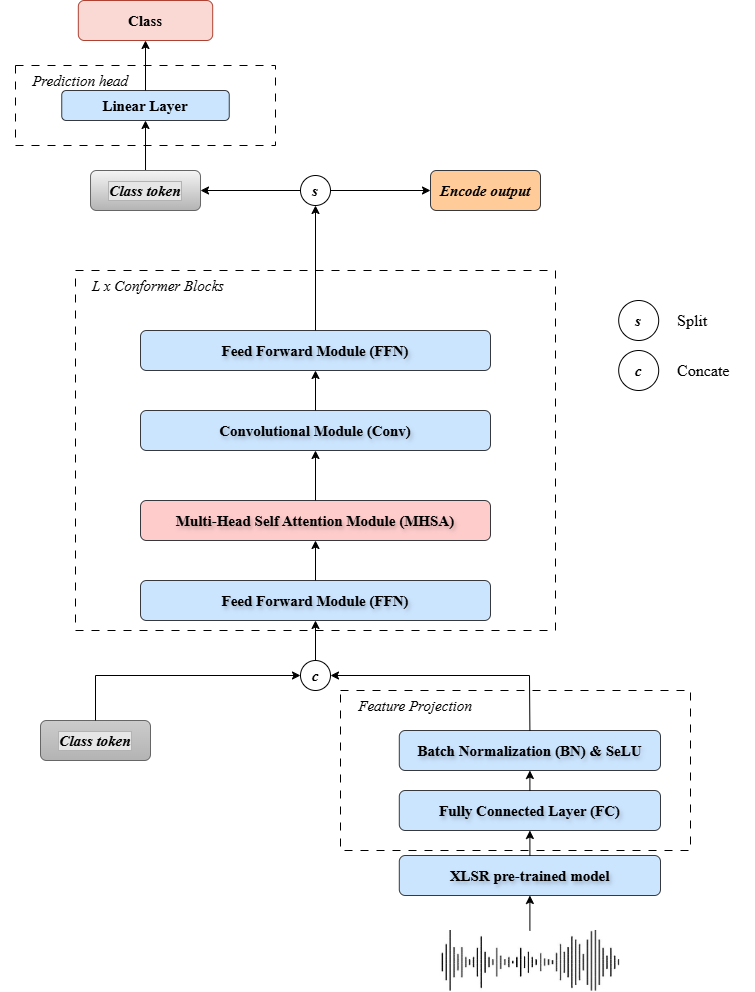}
\end{center}
   \caption{The overall architecture of XLSR-Conformer.}
\label{fig:xlsr}
\end{figure}

We adopt the XLSR-Conformer architectures \cite{rosello23_interspeech} as our baseline. As illustrated in Fig \ref{fig:xlsr}, the baseline model is comprised of two primary components: (i) the pre-trained XLSR \cite{babu2021xlsrselfsupervisedcrosslingualspeech}, a variant of the wav2vec 2.0 \cite{baevski2020wav2vec20frameworkselfsupervised} model that serves as a feature extractor for capturing contextualized representations from high-dimensional speech signals, and (ii) the Conformer Encoder. For a given input speech signal $O$, the SSL model produces a sequence of $T$-length features represented as $X = \text{SSL}(O) = (x_t \in \mathbb{R}^D \mid t = 1,\ldots, T)$, where $D$ is the output dimension of the SSL model.

Subsequently, the extracted features $X$ are reduced to a lower-dimensional space by passing them through an MLP with a SeLU activation function prior to being input into the Conformer Encoder. These projected features are denoted as $\tilde{X} = \text{SeLU}(\text{Linear}(X))$, where $\tilde{X}=(\tilde{x}_t \in \mathbb{R}^{D'} \mid t = 1,\ldots, T)$.

The Conformer Encoder consists of $L$ stacked Conformer blocks, with each block comprising a Multi-Head Self-Attention (MHSA) mechanism and a Convolutional Module, both placed between two feedforward layers. To adapt the sequence-to-sequence architecture of the Conformer for classification tasks, a learnable classification token is appended to the input embeddings of the Conformer Encoder. Specifically, $\tilde{X}_{in} = [\tilde{X}, \tilde{X}_\text{CLS}], \tilde{X} \in \mathbb{R}^{T \times {D'}}, \tilde{X}_{CLS} \in \mathbb{R}^{1 \times {D'}}$, where $[\cdot, \cdot]$ indicates the concatenation operation. The classification token $\tilde{X}_{CLS}$ is designed to aggregate information throughout the Conformer layers, and its state at the final layer's output is passed through a linear layer for classifying the input speech signal as either bonafide or spoof. During the training process, $\tilde{X}_{CLS}$ learns to encapsulate the most critical features required to differentiate synthetic speech from genuine speech.

%%Truong et al. \cite{truong24b_interspeech} proposed integrating the channel representation head token into the temporal input token within the MHSA, called XLSR-Conformer+TCM model. Therefore, the model is capable of learning the temporal-channel dependencies from the input sequence. The architecture is similar to the XLSR-Conformer baseline, with the modifications only represented in the MHSA module. The XLSR-Conformer+TCM reduces the deepfake detection error rate by 26\% relatively while being competitive on the logical access benchmark of ASVspoof 2021.

In the baseline architectures mentioned above, the use of MLPs is hypothesized to result in suboptimal performance due to the challenges posed by high-dimensional contextualized representations. To address this, we propose replacing the MLP with a straightforward ChebyKAN layer, which efficiently approximates the functional mapping while maintaining computational efficiency. Furthermore, we introduce the Kanformer architecture, which integrates Conformer and ChebyKAN, aiming to enhance the learning capability of SSL features.

\subsection{Kanformer Block Architectures}
\label{ssec:kanformer}

\begin{figure*}
\begin{center}
\includegraphics[width=0.8\linewidth]{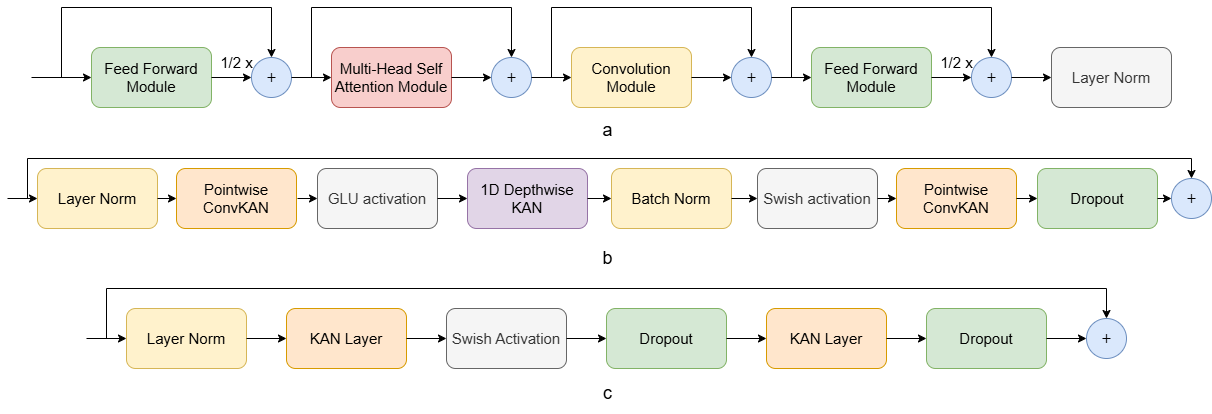}
\end{center}
   \caption{The novel architecture of: a) Kanformer Block; b) Kanformer Feed Forward Module; c) Kanformer Convolution Module.}
\label{fig:kanformer}
\end{figure*}

Kanformer is inspired by a Conformer block, which consists of four sequentially arranged modules: an initial feed-forward module, a self-attention module, a convolution module, and a concluding feed-forward module. Our proposed Kanformer alters the architectures of the feed-forward module and the convolution module - relative to the standard Conformer architecture, as shown in Fig \ref{fig:kanformer}a. 

\subsubsection{Kanformer Feed Forward Module}
The feed forward module as proposed in \cite{gulati2020conformerconvolutionaugmentedtransformerspeech} consists of a Swish activation between two linear transformation and a nonlinear activation. \cite{gulati2020conformerconvolutionaugmentedtransformerspeech} also applies layer normalization, dropout and residual connections, which facilitates the network's regularization process. In this paper, we replace the two linear transformations in the feed-forward module of the Conformer with KAN layers, as illustrated in \ref{fig:kanformer}c. This modification leverages the superior function approximation capabilities of KAN, allowing the network to model complex mappings more effectively. 

\subsubsection{Kanformer Convolution Module}

\cite{gulati2020conformerconvolutionaugmentedtransformerspeech} deploys a convolution module before multi-head self attention layer, which starts with a pointwise convolution and intergrates with a gated linear unit (GLU). Next, a single 1-D depthwise convolution layer is applied. To facilitate the training of deep models, batch normalization is utilized immediately after the convolution. In this paper, we introduce a novel architecture described in \ref{fig:kanformer}b that enhances the conventional Conformer model by replacing its standard convolution layers - specifically, pointwise and depthwise convolutions - with innovative convolution modules integrated with the KAN framework \cite{bodner2024convolutionalkolmogorovarnoldnetworks}. This re-engineering leverages the learnable non-linear activation functions inherent to KAN, enabling more effective approximation of complex functions and improved modeling of high-dimensional speech representations. Moreover, the new design facilitates continual learning during fine-tuning with SSL features, leading to enhanced performance and robustness in synthetic speech detection tasks.

%------------------------------------------------------------------------- 
\section{Experiments}
\label{sec:experiments}
\begin{table*}[h]
\centering
\renewcommand{\arraystretch}{1.1}
\label{tab:table1}
\begin{tabular}{l|cccc|cc}
\hline
{\textbf{Model}} & \multicolumn{2}{c}{\textbf{21LA (Fix)}} & \multicolumn{2}{c|}{\textbf{21LA (Var)}} & \textbf{21DF (Fix)} & \textbf{21DF (Var)}\\ \cline{2-7} 
 & \textbf{EER (\%)} & \textbf{min t-DCF} & \textbf{EER (\%)} & \textbf{min t-DCF} & \textbf{EER (\%)} & \textbf{EER (\%)}\\
 \hline
RawNet2 \cite{9414234} & 9.50 & 0.4257 & - & - & 22.38 & -\\ 
AASIST \cite{jung2021aasistaudioantispoofingusing} & 5.59 & 0.3398 & - & - & - & -\\ 
RawFormer \cite{10096278} & 4.98 & 0.3186 & 4.53 & 0.3088 & - & - \\ 
XLSR-AASIST \cite{tak2022automaticspeakerverificationspoofing} & \textbf{1.00} & \textbf{0.2120} & - & - & 3.69 & -\\ 
\hline
XLSR-Conformer \cite{rosello23_interspeech} (Baseline 1)$\dagger$  & 1.40 &  0.2226  & 1.26 & 0.2200 & 2.79 & 2.98\\ 

XLSR-Conformer + TCM \cite{truong24b_interspeech} (Baseline 2)$\dagger$ & 1.74 & 0.2333 & 2.13 & 0.2445 & 2.74 & 2.77\\ 

XLSR-Kanformer (Proposed) & 1.29 & 0.2173  & \textbf{1.17}  & \textbf{0.2168}  & \textbf{2.36}  & \textbf{2.42} \\ 
%XLSR-GRKAN-Conformer + TCM (Proposed) & \textbf{0.80} & \textbf{0.2067} & \textbf{0.70} & \textbf{0.2042} & 2.54 & 2.62\\ 
\hline

\end{tabular}
\caption{Performance comparison with SOTA models on the ASVspoof 2021 LA and DF evaluation set using fixed-length (Fix) and variable-length (Var) utterance evaluation. The best result is bolded, dash denotes the results are unavailable, $\dagger$ denotes reproduced results.}
\label{tab:table1}
\end{table*}
\subsection{Datasets and Evaluation metrics}
\label{ssec:dataset}

The training and development datasets for our experiments are drawn from the ASVspoof 2019 logical access (LA) track \cite{Todisco2019ASVspoof2F}, which includes both genuine speech and synthetic speech produced by voice conversion and text-to-speech techniques. For evaluation, we utilize the ASVspoof 2021 corpus \cite{10155166}, comprising both the logical access (LA) and deepfake (DF) subsets. The ASVspoof 2021 LA evaluation set contains two known and eleven unknown attack types, while the DF set introduces two additional source datasets. To emulate real-world conditions, the speech data is subjected to various codec and compression alterations. Our primary performance metrics are the equal error rate (EER) and the minimum normalized tandem detection cost function (min t-DCF).

% Besides, we also evaluated our baseline and proposed models on the most recent benchmark in synthetic speech detection to assess the performance on realistic unseen data. The additional evaluation sets include: (i) the In-the-wild \cite{} comprising 20.7 hours of bonafide and 17.2 hours of spoofed audio collected in the wild; and (ii) the hidden subset of ASVspoof 2021 with non-speech segments removed from the utterances, to evaluate the dependence of synthetic speech detection models upon certain data characteristics.

\subsection{Experimental Setup}
\label{ssec:setup}

To facilitate an unbiased comparison, we retained all the configurations outlined in the baseline models \cite{rosello23_interspeech}. For training, every audio sample was standardized to roughly 4 seconds by trimming or padding as necessary. We employed the Adam optimizer with a learning rate of \(10^{-6}\) and applied a weight decay of \(10^{-4}\). The final results were obtained by averaging the metrics of the top 5 models based on their performance on the validation set. Furthermore, early stopping was activated if the validation loss did not improve over 7 consecutive epochs.
%We used the codebase in the baseline model and the official implementation of GR-KAN\footnote{https://github.com/Adamdad/kat.git}.

Furthermore, we enhanced the training data using the RawBoost augmentation technique. In our experiments, we adhered to the same RawBoost configuration and parameter settings as outlined in our baseline study \cite{9746213, truong24b_interspeech}. For evaluation on the LA and DF tracks, we developed two separate SSD systems with tailored RawBoost settings: for the LA track, we combined linear and nonlinear convolutive noise with impulsive, signal-dependent additive noise, whereas for the DF track, we utilized stationary, signal-independent additive noise alongside randomly colored noise.

% \subsection{Evaluation Metrics}

\subsection{Experimental Results}
\label{ssec:result}

% The performance of the proposed model with our reproduced baseline model XLSR-Conformer with TCM and various existing competitive models on the ASVspoof2021 LA and DF evaluation dataset is described in Table \ref{tab:table1}. For fixed-length input evaluation on the LA track, our model achieves an approximately 54\% relative improvement in EER compared to the baseline XLSR-Conformer with TCM (0.80\% vs 1.74\%). The integration of GR-KAN with the XLSR-Conformer not only establishes a new state-of-the-art performance with EERs of 0.80\% and 0.70\% on the ASVspoof LA track but also demonstrates superior performance over other models on the DF track. Notably, our model surpasses the reproduced baseline results by 7.3\% for fixed-length input and 5.42\% for variable-length input.

Table \ref{tab:table1} presents the evaluation results on the ASVSpoof21 LA and DF datasets. Our proposed XLSR-Kanformer models consistently outperform the corresponding baseline systems across all evaluation settings. Specifically, when compared to the XLSR-Conformer baseline (Baseline 1), XLSR-Kanformer demonstrates a relative EER improvement of 7.86\% on the 21LA (Fix) subset (improving from 1.29 to 1.40) and 7.14\% on the 21LA (Var) subset (from 1.26 to 1.17). In addition, the min t-DCF metric shows enhanced calibration, decreasing from 0.2226 to 0.2173. Moreover, when evaluated against the XLSR-Conformer + TCM baseline (Baseline 2), our XLSR-Kanformer further reduces the EER by 25.9\% for 21LA (Fix) and by 45.1\% for 21LA (Var).

For the 21DF task, XLSR-Kanformer outperforms Baseline 1 in both fixed-length and variable-length configurations. In the fixed-length setting, the EER decreases from 2.79 to 2.36, representing a relative reduction of 15.4\%, while in the variable-length setting, it drops from 2.98 to 2.42, an 18.8\% relative reduction. Additionally, when compared with Baseline 2, our model achieves further improvements, with a 13.9\% relative reduction in EER for the fixed-length configuration and a 12.6\% reduction for the variable-length configuration. These results confirm that incorporating ChebyKAN into the Conformer-based system substantially enhances performance compared to the standard MLP approach.

%Table \ref{tab:table2} presents a comparison between our model and the baseline model on both LA and DF tracks, using variable length data for both training and evaluation. Specifically, our model achieves a 59.7\% relative reduction in EER for 21LA and a 13.7\% improvement in min t-DCF (0.2387 → 0.2061), while maintaining comparable performance on 21DF (2.56\% $\rightarrow$ 2.54\%). 

% Our method consistently outperforms the baseline model across various evaluation conditions, demonstrating that KANs can be effectively applied to speech processing tasks, particularly in the context of SSD tasks.

\subsection{Ablation Study and Analysis}
\label{sec:ablation}

\begin{table}[h]
\centering

\begin{tabular}{l|lrr}
\hline
\textbf{SSL Model}                       & \textbf{Projector} & \multicolumn{1}{c}{\textbf{21LA (Fix)}} & \textbf{21DF (Fix)} \\ \hline
\multirow{2}{*}{WavLM Large\tablefootnote{https://huggingface.co/microsoft/wavlm-large}}                   & Conformer                 & 4.07                                       & 11.25                   \\
                                         & Kanformer              & 3.14                                    & 12.29                \\
\multirow{2}{*}{XLSR\tablefootnote{https://huggingface.co/facebook/wav2vec2-xls-r-300m}}                    & Conformer                 & 1.40                                    & 2.79                \\
                                         & Kanformer              & 1.29                                    & 2.36               \\
\multirow{2}{*}{UniSpeech-SAT\tablefootnote{https://huggingface.co/microsoft/unispeech-sat-base-plus}} & Conformer                 & 17.46                                   & 28.46               \\
                                         & Kanformer              & 12.74                                   & 22.98               \\
\multirow{2}{*}{mHuBERT-147\tablefootnote{https://huggingface.co/utter-project/mHuBERT-147}}             & Conformer                 & 21.20                                   & 16.48               \\
                                         & Kanformer              & 16.05                                   & 20.42               \\ \hline
\end{tabular}%
\label{tab:table2}
\caption{EER (\%) result with different SSL models}

\end{table}
In this section, we assess the robustness of our proposed replacement of Conformer with Kanformer in working with features from various different SSL models of various sizes and architectures. We consider: (i) WavLM, a self-supervised model optimized for speech processing tasks; (ii) XLSR, a cross-lingual variant of wav2vec 2.0 designed for multilingual speech representation; (iii) UniSpeech-SAT, which incorporates speaker-aware training to enhance speaker and content modeling; and (iv) mHuBERT-147, a multilingual version of HuBERT trained on 147 languages for robust speech representation. 

Table 2 highlights the impact of replacing Conformer with Kanformer across different SSL models. For the 21LA task, Kanformer consistently reduces the EER across all models, with notable relative improvements of 27.0\% for UniSpeech-SAT and 24.3\% for mHuBERT-147. However, for the 21DF task, while Kanformer improves performance for XLSR, UniSpeech-SAT, and mHuBERT-147, it does not yield gains for WavLM Large, where the EER increases slightly from 11.25 to 12.29. This suggests that while Kanformer enhances robustness in most cases, its effectiveness may vary depending on the SSL model and the characteristics of the evaluation set.

\begin{table}[h]
    \centering
    \begin{tabular}{l|c|c}
        \hline
        & \textbf{LA21}  & \textbf{DF21}\\
        \hline
        XLSR-Conformer w/o TCM (baseline 1) & 1.40 & 2.79 \\
        XLSR-Conformer w TCM (baseline 2) & 1.74 & 2.74 \\
        \textbf{XLSR-Kanformer} & \textbf{1.29} & \textbf{2.36} \\
        \hline
        w/o KAN in Feature Projection & 1.4 & 4.2 \\
        w/o KAN in Feed Forward Module & 1.61 & 2.99 \\
        w/o KAN in Convolution Module & 1.59 & 4.38 \\
        \hline
    \end{tabular}
    \caption{Ablation study of each component in our proposed model on ASV2021 evaluation set (EER).}
    \label{tab:table3}
\end{table}

Table \ref{tab:table3} presents an ablation study assessing the impact of KAN in different components of our proposed XLSR-Kanformer model on the ASV2021 evaluation set. The full XLSR-Kanformer achieves the best performance, with EER reductions compared to both baselines. Removing KAN from the feature projection, feed-forward, or convolution modules leads to performance degradation, particularly in the convolution module, where EER increases significantly (from 1.29\% to 1.59\% on LA21 and from 2.36\% to 4.38\% on DF21). These results highlight the effectiveness of integrating KAN, especially in convolution operations, in improving anti-spoofing performance.

% Meanwhile, Table \ref{tab:table3} compares the performance of both models using Conformer encoders of different sizes. It is evident that regardless of the Conformer type, our model consistently outperforms the baseline \cite{truong24b_interspeech}, further reinforcing the effectiveness of GR-KAN compared to MLP.

% \begin{table}[h]
% \centering
% \renewcommand{\arraystretch}{1.3}
% \caption{Performance comparison with baseline models using different numbers of attention heads and embedding sizes, where ``-B'' corresponds to the Conformer encoder's size of 256 with 4 attention heads, and ``-L'' denotes 512 with 8 attention heads.}
% \label{tab:table3}
% \begin{tabular}{l|c c|c}
% \hline
% {\textbf{System}} & \multicolumn{2}{c|}{\textbf{21LA}} & \textbf{21DF} \\ \cline{2-4} 
%  & \textbf{EER (\%)} & \textbf{min t-DCF} & \textbf{EER (\%)} \\ \hline
% Baseline 2 \cite{truong24b_interspeech} -B & 1.28 & 0.2191 & 3.30 \\ 
% Our model -B & \textbf{1.03} & \textbf{0.2131} & \textbf{2.09} \\ \hline
% Baseline 2 \cite{truong24b_interspeech} -L & 1.73 & 0.2297 & 2.47 \\ 
% Our model -L & \textbf{1.33} & \textbf{0.2213} & \textbf{2.05} \\ \hline
% \end{tabular}
% \end{table}

%------------------------------------------------------------------------ 
\section{Conclusion}
\label{sec:conclusion}

In this work, we introduced a novel integration of the Kolmogorov-Arnold Network (KAN) into the XLSR-Conformer framework, replacing MLP components with KAN and redesigning the Conformer module into Kanformer by incorporating KAN-based architectures. Our proposed XLSR-Kanformer consistently outperforms the baseline XLSR-Conformer and other competitive systems across various evaluation settings, achieving substantial reductions in EER and improvements in min t-DCF. These results highlight the effectiveness of KAN in enhancing both feature extraction and model robustness, particularly in challenging deepfake detection tasks. The demonstrated performance gains strongly encourage further exploration of KAN-based architectures to enhance the effectiveness of other deep learning models in speech anti-spoofing and beyond.

{\small
\bibliographystyle{ieee}
\bibliography{egbib}

\begin{thebibliography}{10}\itemsep=-1pt

\bibitem{babu2021xlsrselfsupervisedcrosslingualspeech}
A.~Babu, C.~Wang, A.~Tjandra, K.~Lakhotia, Q.~Xu, N.~Goyal, K.~Singh, P.~von Platen, Y.~Saraf, J.~Pino, A.~Baevski, A.~Conneau, and M.~Auli.
\newblock Xls-r: Self-supervised cross-lingual speech representation learning at scale, 2021.

\bibitem{baevski2020wav2vec20frameworkselfsupervised}
A.~Baevski, H.~Zhou, A.~Mohamed, and M.~Auli.
\newblock wav2vec 2.0: A framework for self-supervised learning of speech representations, 2020.

\bibitem{bodner2024convolutionalkolmogorovarnoldnetworks}
A.~D. Bodner, A.~S. Tepsich, J.~N. Spolski, and S.~Pourteau.
\newblock Convolutional kolmogorov-arnold networks, 2024.

\bibitem{conneau2020unsupervisedcrosslingualrepresentationlearning}
A.~Conneau, A.~Baevski, R.~Collobert, A.~Mohamed, and M.~Auli.
\newblock Unsupervised cross-lingual representation learning for speech recognition, 2020.

\bibitem{gulati2020conformerconvolutionaugmentedtransformerspeech}
A.~Gulati, J.~Qin, C.-C. Chiu, N.~Parmar, Y.~Zhang, J.~Yu, W.~Han, S.~Wang, Z.~Zhang, Y.~Wu, and R.~Pang.
\newblock Conformer: Convolution-augmented transformer for speech recognition, 2020.

\bibitem{hornik1989multilayer}
K.~Hornik, M.~Stinchcombe, and H.~White.
\newblock Multilayer feedforward networks are universal approximators.
\newblock {\em Neural networks}, 2(5):359--366, 1989.

\bibitem{10096278}
X.~Liu, M.~Liu, L.~Wang, K.~A. Lee, H.~Zhang, and J.~Dang.
\newblock Leveraging positional-related local-global dependency for synthetic speech detection.
\newblock In {\em ICASSP 2023 - 2023 IEEE International Conference on Acoustics, Speech and Signal Processing (ICASSP)}, pages 1--5, 2023.

\bibitem{10155166}
X.~Liu, X.~Wang, M.~Sahidullah, J.~Patino, H.~Delgado, T.~Kinnunen, M.~Todisco, J.~Yamagishi, N.~Evans, A.~Nautsch, and K.~A. Lee.
\newblock Asvspoof 2021: Towards spoofed and deepfake speech detection in the wild.
\newblock {\em IEEE/ACM Transactions on Audio, Speech, and Language Processing}, 31:2507--2522, 2023.

\bibitem{liu2024kankolmogorovarnoldnetworks}
Z.~Liu, Y.~Wang, S.~Vaidya, F.~Ruehle, J.~Halverson, M.~Soljačić, T.~Y. Hou, and M.~Tegmark.
\newblock Kan: Kolmogorov-arnold networks, 2024.

\bibitem{rosello23_interspeech}
E.~Rosello, A.~Gomez-Alanis, A.~M. Gomez, and A.~Peinado.
\newblock A conformer-based classifier for variable-length utterance processing in anti-spoofing.
\newblock In {\em Interspeech 2023}, pages 5281--5285, 2023.

\bibitem{SCHMIDTHIEBER2021119}
J.~Schmidt-Hieber.
\newblock The kolmogorov–arnold representation theorem revisited.
\newblock {\em Neural Networks}, 137:119--126, 2021.

\bibitem{8461375}
D.~Snyder, D.~Garcia-Romero, G.~Sell, D.~Povey, and S.~Khudanpur.
\newblock X-vectors: Robust dnn embeddings for speaker recognition.
\newblock In {\em 2018 IEEE International Conference on Acoustics, Speech and Signal Processing (ICASSP)}, pages 5329--5333, 2018.

\bibitem{ss2024chebyshevpolynomialbasedkolmogorovarnoldnetworks}
S.~SS, K.~AR, G.~R, and A.~KP.
\newblock Chebyshev polynomial-based kolmogorov-arnold networks: An efficient architecture for nonlinear function approximation, 2024.

\bibitem{9746213}
H.~Tak, M.~Kamble, J.~Patino, M.~Todisco, and N.~Evans.
\newblock Rawboost: A raw data boosting and augmentation method applied to automatic speaker verification anti-spoofing.
\newblock In {\em ICASSP 2022 - 2022 IEEE International Conference on Acoustics, Speech and Signal Processing (ICASSP)}, pages 6382--6386, 2022.

\bibitem{9414234}
H.~Tak, J.~Patino, M.~Todisco, A.~Nautsch, N.~Evans, and A.~Larcher.
\newblock End-to-end anti-spoofing with rawnet2.
\newblock In {\em ICASSP 2021 - 2021 IEEE International Conference on Acoustics, Speech and Signal Processing (ICASSP)}, pages 6369--6373, 2021.

\bibitem{tak2022automaticspeakerverificationspoofing}
H.~Tak, M.~Todisco, X.~Wang, J.~weon Jung, J.~Yamagishi, and N.~Evans.
\newblock Automatic speaker verification spoofing and deepfake detection using wav2vec 2.0 and data augmentation, 2022.

\bibitem{Todisco2019ASVspoof2F}
M.~Todisco, X.~Wang, V.~Vestman, M.~Sahidullah, H.~Delgado, A.~Nautsch, J.~Yamagishi, N.~W.~D. Evans, T.~H. Kinnunen, and K.~A. LEE.
\newblock Asvspoof 2019: Future horizons in spoofed and fake audio detection.
\newblock In {\em Interspeech}, 2019.

\bibitem{truong24b_interspeech}
D.-T. Truong, R.~Tao, T.~Nguyen, H.-T. Luong, K.~A. Lee, and E.~S. Chng.
\newblock Temporal-channel modeling in multi-head self-attention for synthetic speech detection.
\newblock In {\em Interspeech 2024}, pages 537--541, 2024.

\bibitem{Vaessen_2022}
N.~Vaessen and D.~A. Van~Leeuwen.
\newblock Fine-tuning wav2vec2 for speaker recognition.
\newblock In {\em ICASSP 2022 - 2022 IEEE International Conference on Acoustics, Speech and Signal Processing (ICASSP)}, page 7967–7971. IEEE, May 2022.

\bibitem{jung2021aasistaudioantispoofingusing}
J.~weon Jung, H.-S. Heo, H.~Tak, H.~jin Shim, J.~S. Chung, B.-J. Lee, H.-J. Yu, and N.~Evans.
\newblock Aasist: Audio anti-spoofing using integrated spectro-temporal graph attention networks, 2021.

\bibitem{zhang2022mfaconformermultiscalefeatureaggregation}
Y.~Zhang, Z.~Lv, H.~Wu, S.~Zhang, P.~Hu, Z.~Wu, H.~yi~Lee, and H.~Meng.
\newblock Mfa-conformer: Multi-scale feature aggregation conformer for automatic speaker verification, 2022.

\end{thebibliography}
}

\end{document}